\documentclass[12pt]{article}
\usepackage[a4paper,left=1.5748in,right=0.59055in,top=1.5748in]{geometry}
\usepackage{amssymb}
\usepackage{setspace}
\usepackage{amsmath}
\usepackage{enumerate}

\newtheorem{thm}{Theorem}[section]
\newtheorem{prop}[thm]{\textsc{Proposition}}
\newtheorem{lem}[thm]{\textsc{Lemma}}
\newtheorem{cor}[thm]{\textsc{Corollary}}

\newcommand{\pf}{\noindent {\bf Proof.  }}
\newcommand{\qed}{\hfill $\Box$ \\}

\def\0{{\mathbf 0}}
\def\C{{\mathcal C}}

\newcommand{\F}{\mathbb{F}}
\newcommand{\Z}{\mathbb{Z}}

\begin{document}

\title{Long quasi-polycyclic $t-$CIS codes} \author{Adel Alahmadi\thanks{Math. Dept., King Abdulaziz University, Jeddah, Saudi Arabia, {Email: \tt adelnife2@yahoo.com}}, Cem G\"{u}neri\thanks{Sabanc\i \ University, FENS, 34956 Istanbul, Turkey, {Email: \tt guneri@sabanciuniv.edu}}, Hatoon Shoaib\thanks{Math. Dept., King Abdulaziz University, Jeddah, Saudi Arabia, {Email: \tt hashoaib@kau.edu.sa }}, Patrick Sol\'e\thanks{CNRS/LAGA, Universit\'e de Paris 8, 93 526 Saint-Denis, France, {Email: \tt sole@math.univ-paris13.fr}} } \date{} \maketitle

\begin{abstract}
We study complementary information set codes of length $tn$ and dimension $n$ of order $t$ called ($t-$CIS code for short). Quasi-cyclic and
quasi-twisted $t$-CIS codes  are enumerated by using their concatenated structure. Asymptotic existence results are derived for one-generator
and have co-index $n$ by Artin's conjecture for quasi cyclic and special case for quasi twisted. This shows that there are infinite families of
 long QC and QT $t$-CIS codes with relative distance satisfying a modified Varshamov-Gilbert bound for rate $1/t$ codes.
 Similar results are defined for the new and more general class of quasi-polycyclic codes introduced recently by Berger and Amrani.
 \end{abstract}

\vspace*{1cm}


\section{Introduction}

In \cite{CGKS} a new class of rate one-half binary codes is introduced: complementary information set codes. A binary linear code of length $2n$
and dimension $n$ is called a complementary information set code (CIS code for short) if it has two disjoint information sets. The motivation
was Booolean masking, a countermeasure aimed at avoiding information leak in cryptographic computations made in embarked electronics.

In this paper, we consider $q$-ary codes of rate $1/t$ that admit $t\ge 2$ pairwise disjoint information sets. These codes are called
complementary information set codes of order $t$ ($t-$CIS code for short). These latter codes were introduced and studied for $q=2$ in
\cite{CFGKKS}, where asymptotic existence results are derived for long linear CIS codes. In the present paper, we derive similar results for
three algebraic classes of codes: quasi-cyclic (QC) codes, quasi-twisted (QT) codes and the more recent class of quasi-polycyclic (QPC) codes,
introduced in \cite{BA}, which contains the first two as subclasses. We also give some numerical examples in modest lengths. Since binary codes
are more important for hardware implementations, we describe a process to derive binary CIS codes from $2^m$-ary one for any integer $m>1.$

The material is organized as follows. The next section collects the necessary notions and notations needed in the rest of the paper, as well as the process just mentioned.  Section 3
gives some examples with optimum distance in modest lengths. Section 4 contains exact enumeration formulae. Section 5 builds on Section 4 to
study the asymptotic performance of QC, QT, and QPC $t$-CIS codes. Section 6 extends these results to $\Z_4$-codes. Section 7 puts these results
into perspective, and points out some challenging open problems.

\section{Definitions and Background}\label{background}
A code $C$ of length $tn$ which has $t$ pairwise disjoint information sets is called a $t$-complementary information set ($t$-CIS) code. When
these sets are made from consecutive integers, we call this partition the natural partition. {\bf We will make this assumption throughout the
article for the CIS codes we study.} These codes were introduced in \cite{CGKS} for $t=2$ and later generalized in \cite{CFGKKS} to higher
$t$'s. Let us note that both articles study only binary codes.

We will assume throughout that the $t$-CIS code $C$ dealt with is in standard form. Hence, if the alphabet of $C$ is $\mathcal{A}$, then $C$ can
be described as
$$C=\{(u,F_1(u),F_2(u),\ldots ,F_{t-1}(u)): u\in \mathcal{A}^n \},$$
where $F_i$'s are permutations of $\mathcal{A}^n$. Let us note that CIS codes can be studied both in linear and nonlinear cases. For a $q$-ary
linear $t$-CIS code, $\mathcal{A}=\F_q$ and $F_i$'s are $\F_q$-linear isomorphisms whereas for a $\Z_4$ CIS code for instance, which is studied
in \cite{CFGKKS,CGKS}, $\mathcal{A}=\Z_4$ and $F_i$'s are simply permutations of $\Z_4^n$.

Note that for an $\F_q$-linear $t$-CIS code $C$, the following is a generating matrix in standard form
$$G=\left(I_n:M_1: \cdots :M_{t-1} \right),$$
where $M_1,\ldots , M_{t-1}$ are $n\times n$ invertible matrices corresponding to the $\F_q$-linear isomorphisms $F_1,\ldots ,F_{t-1}$.

We will study CIS codes coming from codes over rings (i.e. $R$-submodule of $R^t$, where $R$ is a ring). The rings we analyze have the form
$$R=\F_q[x]/\langle h(x) \rangle,$$
where $h(x)$ is some monic polynomial of degree $n$ in $\F_q[x]$. The codes we study will be generated by one element as modules. In particular,
the generator of the code $C$ will be of the form
\begin{equation}\label{generator}
(1,a_1(x),a_2(x),\ldots ,a_{t-1}(x))\in R^t,
\end{equation}
which yields a corresponding generating matrix of size $n\times tn$ in systematic form
\begin{equation}\label{gen-matrix}
G=\left(I_n:A_1: \cdots :A_{t-1} \right),
\end{equation}
where for all $i$, the matrix $A_i$ is a ``circulant" matrix of size $n\times n$ corresponding to $a_i(x)$. More specifically, this means that
the first row of $A_i$ is the $x$-expansion of $a_i(x)$ (i.e. simply coefficients) and the following rows are the $x$-expansions of
$xa_i(x),x^2a_i(x),\ldots$ mod $h(x)$.

For a general polynomial $h(x)\in \F_q[x]$, we will call the codes in $R^t$ quasi-polycyclic (QPC). This kind of codes have been recently
studied in \cite{BA} (see also \cite{ADLS}). If $h(x)=x^n-1$, then the related codes are quasi-cyclic (QC) codes of index $t$ (see \cite{LS})
and $A_i$'s are circulant matrices in the usual sense. If $h(x)=x^n-\alpha$ for some $\alpha \not= 1$, then the codes are called quasi-twisted
(QT) of index $t$ (see \cite{J}). The case $t=2$ and $\alpha=-1$ amounts to double negacirculant codes (\cite{AGOSS}). {\bf We will assume for
simplicity that $h(x)$ is a separable (i.e. without repeated roots) polynomial in the QPC case.} This assumption is easily made by saying that
$n$ is relatively prime to $q$ in the QC and QT cases.

Whether such one-generator code families yield a CIS code (i.e. $A_i$'s in (\ref{gen-matrix}) are all invertible) can be characterized by the
polynomials $a_i(x)$ in (\ref{generator}). The following result is stated for QC codes in \cite[Proposition 9.1]{CFGKKS}. We provide the general
statement and its proof for completeness.

\begin{prop}\label{CIS condition}
Let $R=\F_q[x]/\langle h(x) \rangle$ for a monic polynomial $h$ of degree $n$. The code $C$ in $R^t$ generated by one element as in
(\ref{generator}) is $t$-CIS if and only if $\gcd(a_i(x),h(x))=1$ for all $1\leq i \leq t-1$.
\end{prop}

\pf It suffices to show that each matrix $A_i$ in (\ref{gen-matrix}) is invertible, which amounts to showing that the rows coming from
$a_i(x),xa_i(x),\ldots , x^{n-1}a_i(x)$ mod $h(x)$ are linearly independent over $\F_q$. Being coprime to $h$ is equivalent to saying that
$a_i(x)$ is invertible in $R$. Therefore, if there exist $c_0,c_1,\ldots ,c_{n-1} \in \F_q$, not all zero, such that
$$c_0a_i(x)+c_1xa_i(x)+\cdots +c_{n-1}x^{n-1}a_i(x)=0 \mbox{ in $R$},$$
then there exists a nonzero polynomial $c(x)=c_0+c_1x+\cdots +c_{n-1}x^{n-1} \in R$ such that $c(x)a_i(x)=0$, which contradicts invertibility of
$a_i(x)$. \qed

Next we describe the Chinese Remainder Theorem (CRT) decomposition for QPC codes. This has been given in \cite{LS} for QC codes and in \cite{J}
for QT codes. Presentations in these mentioned articles are detailed and pay attention to reciprocals of the irreducible factors of $x^n-1$ and
$x^n-\alpha$. Such care is needed especially for settling the precise relation to dual codes. Since we do not deal with duality here, we will
have a simpler presentation.

Assume again that $h(x)$ is separable and suppose it factors over $\F_q$ into distinct irreducible polynomials as
\begin{equation}\label{h-poly} h(x)=h_1(x)h_2(x)\cdots h_r(x),\end{equation} where degree of $h_i$ is $d_i\geq 1$ for each $i$. Then by CRT, we have the following ring
isomorphism
$$R\cong \F_{q^{d_1}}\oplus \cdots \oplus \F_{q^{d_r}},$$
and this isomorphism naturally extends to
\begin{equation}\label{CRT decomp}R^t\cong (\F_{q^{d_1}})^t \oplus \cdots \oplus (\F_{q^{d_r}})^t .\end{equation}
Via this isomorphism, one can decompose the QPC code $C=\langle(1,a_1(x),\ldots , a_{t-1}(x))\rangle$ into linear codes $C_i$ of length $t$ over
$\F_{q^{d_i}}$'s (for all $i$), which are called the constituents of $C$. The CRT isomorphism enables us to describe the constituents explicitly
as
\begin{equation}\label{consts} C_i=\mbox{Span}_{\F_{q^{d_i}}}\left\{\left(1,a_1(\xi_i),\ldots , a_{t-1}(\xi_i)\right) \right\}, \ \mbox{for all $1\leq i
\leq r$},\end{equation} where for each $i$, $\xi_i$ is a root of the irreducible polynomial $h_i$ in $\F_{q^{d_i}}$. Therefore for a
one-generator QPC code $C$, each constituent is of length $t$ and of dimension 1 over the related finite field.

For applications, CIS codes over $\F_2$ are more interesting. We will construct CIS codes over arbitrary fields $\F_q$ in this article, but the
following result shows that one can obtain a CIS code over a base field from a CIS code over an extension field.

\begin{prop}\label{descent}
Let $q=2^m$ and suppose $C$ is a $t$-CIS code of length $tk$ and dimension $k$ over $\F_q$. Then there exists a binary $t$-CIS code of length
$mkt$ and dimension $mk$.
\end{prop}

\pf Let us denote $tk$ by $n$. Let $\mathcal{B}=\{e_1,\ldots ,e_m\}$ be a basis of $\F_q$ over $\F_2$ and consider the $\F_2$-isomorphism
$\phi_{\mathcal{B}}$ from $\F_q$ to $\F_2^{m}$, sending $x=x_1e_1+\cdots +x_me_m \in \F_q$ to $(x_1,\ldots ,x_m)\in \F_2^{m}$. This isomorphism
naturally extends to the following $\F_2$-linear isomorphism:
\[\begin{array}{cccc}
\phi_{\mathcal{B}}: & \F_q^n & \longrightarrow & \F_2^n \times \cdots \times \F_2^n \\
& (x^1,\ldots , x^n) & \longmapsto & [(x^1_1,\ldots ,x^n_1): \ldots : (x^1_m,\ldots ,x^n_m)]
\end{array}\]
Then $\phi_{\mathcal{B}}(C)$ is a binary code of length $mn$ and dimension $mk$.

Suppose $\mathcal{I} \subset \{1,\ldots ,n\}$ is an information set for $C$ and let $G_{\mathcal{I}}$ be the $k\times k$ matrix over $\F_q$
whose column indices are determined by $\mathcal{I}$. Note that $G_\mathcal{I}$ is a submatrix of a generating matrix $G$ of $C$. Since
$G_\mathcal{I}$ is of full rank $k$, the image of the subcode generated by $G_\mathcal{I}$ under $\phi_\mathcal{B}$ (i.e.
$\phi_\mathcal{B}(\langle G_\mathcal{I} \rangle )$) is a binary code of length and dimension $mk$ over $\F_2$. If we write the coordinates in
$\F_2^{mn}$ as $\{1,\ldots ,n\}\times \{1,\ldots ,m\}$, then the above discussion shows that $\mathcal{I}\times \{1,\ldots ,m\}$ is an
information set for $\phi_{\mathcal{B}}(C)$. Since $C$ is $t$-CIS, its coordinates have a partition into information sets $\mathcal{I}_1,\ldots
,\mathcal{I}_t$. Discussion above shows that $\mathcal{I}_1\times \{1,\ldots ,m\},\ldots ,\mathcal{I}_t\times \{1,\ldots ,m\}$ is a partition
for the coordinates of $\phi_{\mathcal{B}}(C)$. \qed

\section{Numerical Examples}

\subsection{Quasi cyclic $t-$CIS codes}
\subsubsection{$t=2$}
 Assume that $\C$ is QC 2-CIS codes of length $2n$ and dimension $n$. The search was done in Magma \cite{BCP}, for $q=2$ and random
$a_1(x)$. \\
\begin{center}
\begin{tabular}{|c||c|c|c|c|c|c|c|c|}
  \hline
  \textit{\textbf{$n$}} & 2 & 3& 4 &  5 & 6 & 7 & 8 & 9    \\
  \hline
  \textit{\textbf{$d$}} & 2 & 3 &4 &4 &  4 &  4 &  5 &  6  \\
  \hline
  \textit{\textbf{$d^*$}} & 2 & 3 & 4 & 4 & 4  &  4 &  5 & 6   \\
  \hline
  \textit{\textbf{$d^{**}$}} & 2 & 3 & 4 & 4 & 4  & 4  & 5  &  6  \\
  \hline
\end{tabular}
\end{center}

Here, $\emph{d}$ is the minimum distance for QC 2-CIS codes that we
computed, $\emph{$d^{*}$}$ is the minimum distance for QC 2-CIS codes from \cite{CGKS}, and
 $\emph{$d^{**}$}$ is the highest minimum distance of a linear code of given length and dimension \cite{G}.\\

\subsubsection{$t=3$}
 Assume that $\C$ is QC 3-CIS codes of length $3n$ and dimension $n$. The search was done in Magma \cite{BCP}, for $q=2$ and random
$a_i(x)$ where $i=1,2$. \\
\begin{center}
\begin{tabular}{|c||c|c|c|c|c|c|c|}
  \hline
  \textit{\textbf{$n$}} & 2 & 3& 4 &  5 & 6 & 7 & 8     \\
  \hline
  \textit{\textbf{$d$}} & 4 & 4 & 6 & 7 & 8 & 8 & 8   \\
  \hline
  \textit{\textbf{$d^*$}} & 4 & 4 & 6 & 7 & 8 & 8 & 8    \\
  \hline
  \textit{\textbf{$d^{**}$}} & 4 & 4 & 6 & 7 & 8 & 8 & 8   \\
  \hline
\end{tabular}
\end{center}

Here, $\emph{d}$ is the minimum distance for QC 3-CIS codes that we
computed, $\emph{$d^{*}$}$ is the minimum distance for QC 3-CIS codes from \cite{CFGKKS}, and
 $\emph{$d^{**}$}$ is the highest minimum distance of a linear code of given length and dimension \cite{G}.

\subsection{Quasi twisted $t-$CIS codes}
\subsubsection{$t=2$}
Assume that $\C$ is QT 2-CIS codes of length $2n$ and dimension $n$. The search was done in Magma \cite{BCP}, for $q=4$ and random
$a_1(x)$. \\
\begin{center}
\begin{tabular}{|c||c|c|c|c|c|c|c|}
  \hline
  \textit{\textbf{$n$}} & 2 & 3& 4 &  5 & 6 & 7 & 8     \\
  \hline
  \textit{\textbf{$d$}} & 3 & 3 & 4 & 5 & 5 & 6 & 6     \\
  \hline
  \textit{\textbf{$d^*$}} & 3 & 4 & 4 & 5 & 6 & 6 & 7      \\
  \hline
\end{tabular}
\end{center}
Here, $\emph{d}$ is the minimum distance for QT 2-CIS codes that we computed, and $\emph{$d^{*}$}$ is the highest minimum distance of a linear code of given length and dimension \cite{G}.\\

\subsubsection{$t=3$}
  Assume that $\C$ is QT 3-CIS codes of length $3n$ and dimension $n$. The search was done in Magma \cite{BCP}, for $q=4$ and random
$a_i(x)$ where $i=1,2$. \\
\begin{center}
\begin{tabular}{|c||c|c|c|c|c|c|c|}
  \hline
  \textit{\textbf{$n$}} & 2 & 3& 4 &  5 & 6 & 7 & 8     \\
  \hline
  \textit{\textbf{$d$}} & 4 & 6 & 7 & 8 & 9 & 10 & 11      \\
  \hline
  \textit{\textbf{$d^*$}} & 4 & 6 & 7 & 8 & 10 & 11 & 12        \\
  \hline
\end{tabular}
\end{center}
Here, $\emph{d}$  is the minimum distance for QT 3-CIS codes that we computed, and $\emph{$d^{*}$}$ is the highest minimum distance of a linear code of given length and dimension \cite{G}.

\subsection{Quasi polycyclic $t-$CIS codes}
\subsubsection{$t=2$}
Assume that $\C$ is QPC 2-CIS codes of length $2n$ and dimension $n$. The search was done in Magma \cite{BCP}, for $q=2$ and random
$a_1(x)$. \\
\begin{center}
\begin{tabular}{|c||c|c|c|c|c|c|}
  \hline
  \textit{\textbf{$n$}} & 2 & 3& 4 &  5 & 6 & 7      \\
  \hline
  \textit{\textbf{$d$}} & 2 & 2 & 3 & 3 & 4 & 4       \\
  \hline
  \textit{\textbf{$d^*$}} & 2 & 3 & 4 & 4 & 4 & 4       \\
  \hline
\end{tabular}
\end{center}
Here, $\emph{d}$ is the minimum distance for QPC 2-CIS codes that we computed, and $\emph{$d^{*}$}$ is the highest minimum distance of a linear code of given length and dimension \cite{G}.\\

\subsubsection{$t=3$}
 Assume that $\C$ is QPC 3-CIS codes of length $3n$ and dimension $n$. The search was done in Magma \cite{BCP}, for $q=2$ and random
$a_i(x)$ where $i=1,2$. \\
\begin{center}
\begin{tabular}{|c||c|c|c|c|c|c|}
  \hline
  \textit{\textbf{$n$}} & 2 & 3& 4 &  5 & 6 & 7      \\
  \hline
  \textit{\textbf{$d$}} & 4 &  4 & 5  & 5   & 6  &  6       \\
  \hline
  \textit{\textbf{$d^*$}} & 4 &  4 & 6  & 7   & 8  & 8       \\
  \hline
\end{tabular}
\end{center}
Here, $\emph{d}$ is the minimum distance for QPC 3-CIS codes that we computed, and $\emph{$d^{*}$}$ is the highest minimum distance of a linear code of given length and dimension \cite{G}.

\section{Enumeration}
We prove enumeration results, which will be used for asymptotic analysis in the next section.

\begin{prop}\label{enum3}
Let $h(x)$ be a separable polynomial of degree $n$ in $\F_q[x]$ with irreducible factorization as in (\ref{h-poly}) and consider one-generator
QPC codes in $R^t$. The number of such QPC $t$-CIS codes of length $tn$ is
$$\left(q^{d_1}-1 \right)^{t-1} \cdots \left(q^{d_r}-1 \right)^{t-1}.$$
\end{prop}

\pf By Proposition \ref{CIS condition}, for $C=\langle (1,a_1(x),\ldots ,a_{t-1}(x))\rangle$ to be $t$-CIS, the condition $\gcd(a_j(x),h(x))=1$
holds for all $j$. Therefore $a_j(x)$ does not vanish at any root of $h$ and hence each constituent $C_i$ of $C$ is of dimension one with a
generator $(1,a_1',\ldots , a_{t-1}')$, where $a_j'\in \F_{q^{d_i}}^*$ for all $j$. Hence there are $q^{d_i}-1$ such constituents for each $i$
and the product of all such choices yield all the one-generator $t$-CIS codes. \qed

The following special cases for QC and QT $t$-CIS codes follow immediately from Proposition \ref{enum3}. These consequences will also be needed
in the next section.

\begin{cor}\label{enum1}
Let $n$ be a prime which is relatively prime to $q$. If $x^n-1$ factors as a product of two irreducible polynomials over $\F_q$ as
$x^n-1=(x-1)m(x)$ , then the number of QC $t$-CIS codes over $\F_q$ of length $tn$ and index $t$ is $N= \left((q-1)(q^{n-1}-1)\right)^{t-1}$.
\end{cor}

\begin{cor}\label{enum2}
Let $q$ be a prime power, $n \geq 2$ be an integer and $\alpha \in \F^*_q$. If $x^n-\alpha$ is irreducible in $\F_q[x]$, then the number of the
QT $t$-CIS codes over $\F_q$ of length $tn$ and index $t$ is $N=(q^{n}-1)^{t-1}$.
\end{cor}

\section{Asymptotics}

Our first goal is to show that a class of $t$-CIS QPC codes satisfy a modified Gilbert-Varshamov bound. Recall that $R$ denotes the ring
$\F_q[x]/\langle h(x) \rangle$ for the separable polynomial $h$ of interest. We need some preparation.

\begin{lem}\label{asym1}
Let $h(x)\in \F_q[x]$ be a polynomial of degree $n$ which has $r$ distinct irreducible factors of equal degree.  If $b=(b_0(x),b_1(x),\ldots ,
b_{t-1}(x))\in R^t$ is nonzero, then there exist at most $(q^{n/r}-1)^{(t-1)(r-1)}$ codes $C_{a}=\langle (1,a_1(x),\ldots ,a_{t-1}(x))\rangle
\subset R^t$ which are $t$-CIS QPC and contain $b$.
\end{lem}

\pf By assumption on $h$, $R^t$ is isomorphic to $\F_{q^{n/r}}^t \oplus \cdots \oplus \F_{q^{n/r}}^t$ and the code $C_a$ has $r$ constituents of
dimension 1. Let us denote the image of $(1,a_1(x),\ldots ,a_{t-1}(x))$ under the CRT isomorphism into the $i^{th}$ constituent by
$(1,a_1^i,\ldots ,a_{t-1}^i)$ (for $1\leq i \leq r$). The same notation is valid for the image of $b=(b_0(x),b_1(x),\ldots , b_{t-1}(x))$ under
CRT. Note that $C_a$ being CIS amounts to $a_j^i$ being nonzero in $\F_{q^{n/r}}$ for all $1\leq j \leq t-1$ and $1\leq i \leq r$. Observe that
$b\in C_a$ amounts to following relations:
$$b_j^i=b_0^ia_j^i, \ \mbox{for all $1\leq j \leq t-1$ and $1\leq i \leq r$}.$$
If $b_0^i=0$ for some fixed $i$, then $b_j^i=0$ for all $j$, in which case $a_j^i$'s can be arbitrarily chosen from nonzero elements in
$\F_{q^{n/r}}$ for all $1\leq j \leq t-1$ (i.e. $(q^{n/r}-1)^{t-1}$ choices for $a_1^i,a_2^i,\ldots ,a_{t-1}^i$). Since $b$ is nonzero, at most
$r-1$ of $b_0^i$'s can be zero. When $b_0^i\not= 0$ for a fixed $i$, $a_j^i$'s are uniquely determined by $a_j^i=b_j^i/b_0^i$ for all $1\leq j
\leq t-1$. Hence, the maximum number of possible $a$ choices so that $b\in C_a$ for an arbitrary $b$ is obtained if $b_0^i=0$ for $r-1$ values
of $i$. In this case there exist $(q^{n/r}-1)^{(t-1)(r-1)}$ possible $a$'s containing the given $b$. \qed

The following is also needed for the asymptotic result on a class of $t$-CIS QPC codes.

\begin{lem}\label{poly}
Let $q$ be a prime power and $r$ be a fixed positive integer. There exists $N>0$ such that for infinitely many $n>N$, there exist $r$ distinct
monic irreducible polynomials of degree $n/r$ over $\F_q$.
\end{lem}

\pf The number of monic irreducible polynomials of degree $n/r$ over $\F_q$ is given by
$$\frac{1}{n/r}\sum_{d|\frac{n}{r}} \mu \left(\frac{n}{rd}\right)q^d.$$
Since $q$ and $r$ are fixed, this number is of the order of $rq^{n/r}/n$ for large $n$. So for large enough $N$ and any $n>N$, we have
$r<rq^{n/r}/n$ (or equivalently $n<q^{n/r}$) and therefore there exist $r$ distinct monic irreducibles of equal degree $n/r$. It is clear from
the enumeration formula of irreducible polynomials that for $n'=n+r$, there are more irreducible polynomials of degree $(n+r)/r=n/r+1$, hence
there are $r$ distinct monic irreducibles of degree $n/r+1$. The claim follows inductively. \qed

Recall that the $q$-ary entropy function is defined for $0<y< \frac{q-1}{q}$ by $$ H_q(y)=y\log_q(q-1)-y\log_q(y)-(1-y)\log_q(1-y).$$

\begin{thm} \label{asym2}
Let $q$ be a fixed prime power and $r$ be a fixed positive integer. Let $h(x)\in \F_Q[x]$ be a polynomial which has $r$ distinct monic
irreducible factors $h_1,\ldots ,h_r$ of equal degrees. Let $C(r)$ denote the class of one-generator QPC codes with respect to such $h$. Then
for all $t\geq 2$, there exist infinitely many $t$-CIS QPC codes of rate $1/t$ in $C(r)$, with relative distance $\delta$ satisfying
$$H_q(\delta)\geq \frac{t-1}{rt}.$$
\end{thm}

\pf Note that there exist codes of arbitrarily long lengths in $C(r)$ by Lemma \ref{poly}. The $t-$CIS QPC codes containing a vector of weight
$d\sim t\delta n$ or less are by standard entropic estimates and Lemma \ref{asym1} of the order $(q^{n/r}-1)^{(t-1)(r-1)} \times q^{tn
H_q(\delta)}$, up to subexponential terms. This number will be less than the total number of $t-$CIS QPC codes, which is by Proposition
\ref{enum3} of the order of $(q^{n/r}-1)^{r(t-1)}\sim q^{n(t-1)}$. \qed

Next we specialize to $t$-CIS QT codes. Now, $R=\F_q[x]/\langle x^n-\alpha \rangle$ for some nonzero $\alpha \in \F_q$. We will assume that
$x^n-\alpha$ is irreducible in $\F_q[x]$. By \cite[Theorem 3.75]{LN}, this is true if and only if the following two conditions are satisfied:
\begin{itemize}
  \item[(i)] Each prime factor of $n$ divides the order $a$ of $\alpha$ in $\F_q^*$, but does not divide $(q-1)/a.$
  \item[(ii)] $q\equiv 1$ (mod 4)  if $n\equiv 0$ (mod 4).
\end{itemize}
Hence, there are infinitely many $n$ such that $x^n-\alpha$ is irreducible in $\F_q[x]$.



\begin{thm} \label{asym4}
Let $q$ be a prime power.  For any fixed integer $t\ge 2,$ there are infinite families of $t-$CIS QT codes of rate $1/t$ and of relative
distance $\delta$ satisfying  $H_q(\delta)\geq \frac{t-1}{t}$.
\end{thm}

\pf The infinitude of $n$ such that $x^n-\alpha$ is irreducible in $\F_q[x]$ is guaranteed by the result from \cite{LN} above.
More specifically, if $e$ is an integer dividing $q-1$ such that $(e, \frac{q-1}{e})=1,$ and $p$ is a prime divisor of $e,$ then fixing an $\alpha$ of order $e$
we see that for all $i,$ the binomial $x^{p^i}-\alpha$ is irreducible in $\F_q[x].$
We complete the proof as a special case of $r=1$ of Theorem \ref{asym2}. \qed

We give one more asymptotic result, for $t$-CIS QC codes. We will consider $x^n-1 \in \F_q[x]$ for a prime $n$. In number theory, Artin's
conjecture on primitive roots states that if $q$ is neither a perfect square nor $-1$, then $q$ is a primitive root modulo infinitely many
primes $n$ (\cite{M}). This was proved conditionally under Generalized Riemann Hypothesis by Hooley \cite{H}. In this case, by the
correspondence between cyclotomic cosets and irreducible factors of $x^n-1$ \cite{HP}, the factorization of $x^n-1$ into irreducible polynomials
over $\F_q$ contains exactly two factors, one of which is $x-1$ (\cite{CPW}). Note the difference with QPC and QT cases considered above. So we
need to prove the following claim.

\begin{lem}\label{asym5}
Let $q$ be a prime power and $n$ be a prime number which does not divide $q$. If $b=(b_0(x),b_1(x),\ldots , b_{t-1}(x))\in R^t$ is nonzero and
of weight less than $n$ in each coordinate in $R^t$, then there exists at most one code $C_{a}=\langle (1,a_1(x),\ldots ,a_{t-1}(x))\rangle
\subset R^t$ which is $t$-CIS QC and contains $b$.
\end{lem}

\pf We sketch a proof analogue to the one in \cite{CPW} in the case $t=2$. Denote by $(b'_i,b''_i)$ (resp. $(a'_i,a''_i)$) the image of $b_i(x)$
(resp. $a_i(x)$) in the CRT decomposition of the ring $R.$
If $b \in C_a,$ then, for all $1\le i\le t,$ we have that  $b_i=b_0 a_i.$ Hence, we have $b'_i=b'_0 a'_i,$ and $b''_i=b''_0 a''_i.$
We discuss four cases depending on values of $b_0.$

If both $ b'_0 \neq 0,$ and $ b''_0 \neq 0,$ then $C_a$ is unique.

If $ b'_0 = b''_0= 0,$ then  $b'_i = b''_i= 0,$ and $b=0,$ which is impossible by Hypothesis.

If $ b'_0 \neq 0,$ and $ b''_0=0,$ then $b_i(x)\neq 0$ is a multiple of $(1,x,\dots,x^{n-1})$ hence of weight $n.$

\qed

The proof of the next result follows by Lemma \ref{asym5}, by an argument similar to that of Theorem \ref{asym2}, hence it is omitted.

\begin{thm}\label{asym6}
Let $q$ be a prime power, and $n$ be prime. If $x^n-1=(x-1)u(x)$,with $u$ irreducible, then for any fixed integer $t\ge 2,$ there are infinite families of $t-$CIS QC
codes of index $t$, rate $1/t$ and of relative distance $\delta$, satisfying  $H_q(\delta)\geq \frac{t-1}{t}$.\\
\end{thm}

\section{$\Z_4$-codes}

Quasi-twisted (QT) $\Z_4$-codes have been introduced in \cite{AC}. Consider the ring $R_4(n)=\Z_4[x]/\langle x^n+1 \rangle.$ A QT code of index
$t$ and length $tn$ over $\Z_4$ is an $R_4(n)$-submodule of $R_4(n)^t.$ Note that there is a well-known ring isomorphism between $R_4(n)$ and
the ring of negacirculant matrices of size $n,$ over $\Z_4.$ In particular such a code will be multinegacirculant of index $t$ if its generator
matrix is blocked as $[I,A_1,\cdots,A_{t-1} ],$ with $A_i$'s negacirculant matrices. Further, it will be $t$-CIS if each $A_i$ is invertible.

Assume that over $\F_2$ we have the factorization into
two irreducibles $$x^n+1=(x+1)(x^{n-1}+\cdots+x+1).$$
As is well-known \cite{CPW}, this is equivalent to $n$ being prime and $2$ being primitive modulo $n.$
Since $n$ is odd, by Hensel lifting \cite{HKCSS,W}, we have the factorization
$$x^n-1=(x-1)(x^{n-1}+\cdots+x+1),$$
with $x^{n-1}+\cdots+x+1$ irreducible over $\Z_4.$

This factorization yields the CRT decomposition $R_4(n)=\Z_4\oplus GR(4,n-1),$ where $GR(4,d)$ denotes the Galois ring of characteristic $4$ and size $4^d$ \cite{HKCSS,W}.
Denote by $(a'_i,a''_i)$ the image of $A_i$ in that decomposition.
The CIS-ness condition translates immediately into $a'_i=\pm 1,$ and the $a''_i$'s being units of $GR(4,n-1).$
\begin{thm} \label{asym6}
Assume $2$ to be primitive modulo $n,$ a prime. Then for any fixed integer $t\ge 2,$ and large $n,$ there are infinite families of $t-$CIS
 multinegacirculant $\Z_4$-codes  of length $n,$ rate $1/t$ and relative Lee distance $\delta$ satisfying  $H_2(\delta)\geq \frac{t-1}{t}$.\\

\end{thm}

\pf Note first, that there are infinitely many such $n$'s under Artin's conjecture \cite{M}.
The number of QT $\Z_4$-codes of index $t$ satisfying the said sufficient condition for CISness is easily seen to be $(4^n-2^n)^{t-1},$
since the number of units in $\Z_4$ is $2,$ and the number of units in $GR(4,d),$ for any integer $d,$ is $4^d-2^d$ \cite{HKCSS,W}.
The number of such CIS codes containing a nonzero given vector of $\Z_4^{tn}$ of weight $<d$ is at most one, by an easy analogue of Lemma 5.5 .
The size of the Lee ball of radius $t\delta n$ in $\Z_4^{tn}$ equals, by Gray mapping, that the Hamming ball of radius $2t\delta n$ in $\F_2^{tn}.$ Both quantities are, for fixed $t,$ and large $n,$
asymptotically equivalent, up to subexponential terms to $2^{2tnH_2(\delta)}.$ The result follows by the usual expurgation argument.
\qed

\section{Conclusion} In this paper, motivated by the security of embarked electronics, we have studied a class of combinatorial codes (CIS codes),
from the viewpoint of asymptotic performance. The main tools of our study were the CRT and expurgated random coding. While the class of QC codes required the full force of
Artin's conjecture, the asymptotics of QT codes only required some elementary results on factorization of polynomials over finite fields.

We emphasize the fact that QT codes
are understudied in even characteristic. Our existence results might motivate further consideration of this class of codes in finite lengths. More generally, the new
class of quasipolycyclic codes
deserves further rexploration for a wider variety of polynomials $h,$ beyond the separable case. In a general situation, chain rings will appear in the CRT decomposition, a fact
which might complicate the algebraic analysis of the codes.


\end{document}